\documentclass[a4paper, oneside, twocolumn, notitlepage, 10pt]{extarticle_ecoc}
\usepackage{ecoc}
\usepackage{setspace}
\usepackage{multirow}
\usepackage{subfigure}
\usepackage[font=small,labelfont=bf]{caption}
\usepackage{setspace}
\usepackage[nolist]{acronym}%
\usepackage{color}

\usepackage{fancyhdr}

\fancypagestyle{FirstPage}{
	\lfoot{\vspace{-0.5cm} \hrulefill \\ \small \copyright~2022 IEEE. Personal use of this material is permitted.  Permission from IEEE must be obtained for all other uses, in any current or future media, including reprinting/republishing this material for advertising or promotional purposes, creating new collective works, for resale or redistribution to servers or lists, or reuse of any copyrighted component of this work in other works.} 
	
	\lhead{\vspace{-0.5cm} \small This work has been accepted for publication in European Conference on Optical Communication (ECOC) 2022 18-22th September, Basel, Switzerland \\ \hrulefill}
}

\begin{document}
\selectlanguage{american}    % Standard Language

%-------------------------------------------------- Title -----------------------------------------------------%

\title{Schedulers Synchronization Supporting Ultra Reliable Low Latency Communications (URLLC) in Cloud-RAN over Virtualised Mesh PON }%atency in  A novel two-tier virtualized PON transport to achieve ultra-low application-level latency in virtualized MESH-PON enabled MEC based Cloud-RAN}%

%------------------------------------------------- Authors-----------------------------------------------------%
\author{\vspace{-0.3in}
    Sandip Das\textsuperscript{(1)}, Frank Slyne\textsuperscript{(1)}, Daniel Kilper \textsuperscript{(1)}, Marco Ruffini\textsuperscript{(1)}
}

\maketitle                  % Create title and author

%------------------------------------------ Description of Authors ----------------------------------------------%

\begin{strip}
 \begin{author_descr}

   \textsuperscript{(1)} CONNECT Centre, Trinity College Dublin,
   \textcolor{blue}{\uline{dassa@tcd.ie}, \uline{marco.ruffini@tcd.ie}}
 \end{author_descr}
\vspace{-0.1in}
\end{strip}

\setstretch{1.0}

%-------------------------------------------------- Abstract ---------------------------------------------------------%

\begin{strip}
  \begin{ecoc_abstract}
    We propose a mechanism to support URLLC Open-RAN ultra-low latency over a MESH-PON, serving dense deployment of small cell and MEC nodes in an access network. We show the possibility, under given assumptions, to achieve application-to-application end-to-end latency below 1ms.
  \end{ecoc_abstract}
\vspace*{-0.15in}
\end{strip}
%-------------------------------------------------- Introduction Section -------------------------------------------------------%
\begin{acronym}
	\acro{AR}{Augmented Reality}
	\acro{QoS}{Quality of Service}
	\acro{C-RAN}{Cloud Radio Access Networks}
	\acro{CGS}{Coordinated Grant Scheduling}
	\acro{CU}{Central Unit}
	\acro{CTI}{Coordinated Transport Interface}
	\acro{FBG}{Fiber Bragg Grating}
	\acro{GF}{Grant Free}
	\acro{ITS}{Intelligent Transport Systems}
	\acro{MAIO}{Mixed Analytical Iterative Optimization}
	\acro{MFH}{Mobile Fronthaul}
	\acro{RU}{Radio Unit}
	\acro{BBU}{Baseband Unit}
	\acro{DU}{Distributed Unit}
	\acro{DU/CU}{Distributed Unit/Centralised Unit}
	\acro{DCI}{Downlink Control Information}
	\acro{NR}{New Radio}
	\acro{PON}{Passive Optical Network}
	\acro{vPON}{virtual-PON}
	\acro{ODN}{Optical Distribution Network}
	\acro{TWDM}{Time-Wavelength Division Multiplexing}
	\acro{DBA}{Dynamic Bandwidth Allocation}
	\acro{MEC}{Multi Access Edge Computing}
	\acro{CO}{Central Office}
	\acro{OLT}{Optical Line Terminal}
	\acro{RV}{Random Variable}
	\acro{GC}{Grant Cycle}
	\acro{ONU}{Optical Networking Unit}
	\acro{PLOAM}{Physical Layer Operation and Maintenance}
	\acro{PRB}{Physical Resource Block}
	\acro{eCPRI}{evolved Common Public Radio Interface}
	\acro{BS}{Base Station}
	\acro{SPS}{Semi-Persistent Scheduling}
	\acro{TDMA}{Time Division Multiple Access}
	\acro{TTI}{Transmit Time Interval}
	\acro{VRF}{Variable Rate Fronthaul}
	\acro{vPON}{Virtualized PON}
	\acro{UE}{User Equipment}
	\acro{LLS}{Low Layer Split}
	\acro{vRAN}{Virtualized RAN}
	\acro{WLB}{Wavelength Loop Back}
	\acro{WPF}{Wavelength Pass Filter}
\end{acronym}
\thispagestyle{FirstPage}
\section{Introduction}\label{sec:Intro}
	Support for ultra-low latency applications (i.e., of the order of 1ms or less end-to-end) is one of the key requirements for 5G and beyond in order to support mission-critical applications such as \ac{ITS}, industry 4.0, public safety, including use of \ac{AR} technology \cite{5G-NGMN-Verticals}. \ac{C-RAN}, and \ac{MEC} can help support these requirements by enabling network densification and local data processing and storage. From a networking perspective, when considering end-to-end (i.e., from source to destination at the application level) the latency accumulates in three sections: in the RAN (due to transmission distance and stack processing), in the fronthaul transmission from \ac{RU} to \ac{DU/CU} (due to transmission distance and data scheduling if operating over a PON) and in the transport from the DU/CU towards the MEC node running the application. 
	
	In a traditional RAN, the access latency is the amount of time the \ac{UE} application traffic needs to wait for allocation of uplink resource before transmission, which is generally assigned to the UE via a set of \ac{DCI} messages in 5G \ac{NR}. The latency between buffer status report by the UE and the corresponding uplink resource grant allocation via DCI messages is 4 time slots (e.g., 2 ms for 0.5 ms slot duration \cite{5G-NR_Dahlman-Scheduling}). This is the largest contributing factor to RAN access latency. In order to address this bottleneck, \ac{CGS} (for uplink) and \ac{SPS} (in downlink) were proposed \cite{5G-NR_Dahlman}, which semi-statically pre-allocate uplink resources (typically a group of \acp{PRB}) to UEs so that they can send their uplink traffic without making a request, thus avoiding waiting for the scheduling of uplink resources. 
 	
 	The second source of delay, as mentioned above, is given by the uplink scheduling mechanism, when the RAN is transported over a PON. This latency becomes critical if the RAN works on an eCPRI split (i.e., above split 6, with split 7.2 being typical for Open-RAN \cite{ORAN_FH_Standard}). Here, if the PON and RAN schedulers are not coordinated, data from the UE will also need to queue at the ONU side waiting for the PON upstream grant to be provided by the OLT. This coordination issue was recently solved with the development of cooperative Dynamic Bandwidth Allocation (Co-DBA) \cite{Co-DBA} implemented over a \ac{CTI}\cite{O-RAN-CTI}. This requires the OLT to fetch prior UE uplink scheduling information from DU/CU and use it to estimate the packet size and arrival time at the RU (which in this case is collocated with an ONU). Then the OLT can pre-assign uplink resources to the ONU so that the packets from the UEs have minimal queuing once they arrive at the ONU. 
 	
 	However, this does not solve the issue of latency at the application level. As mentioned above, low latency requires the use of methods like \ac{CGS} in the RAN, where information about incoming data from the UE is not known in advance and thus cannot be passed to the OLT for CTI coordination. %This would incur additional queuing latency with CO-DBA if the allocated \ac{CGS} resource and the traffic over it is under-estimated.
 	CTI currently does not support a RAN that uses the CGS for ultra-low latency. One of the contributions of this work is introducing an updated CTI that can support low-latency RAN operations. %mechanism Therefore, the CTI interface and the corresponding DBA needs to be updated to process this information properly to achieve a ultra-low RU-DU fronthaul transport latency also with RAN access latency significantly reduced via \ac{CGS}. 
 	
 	The third source of network latency is the data transmission between the DU/CU and the server running the application. Typically, this is sent over the network to a \ac{CO} or a \ac{MEC} node, and can involve multiple layer 2 or layer 3 hops, depending on the network configuration. We address these shortcomings through the new MESH-PON approach described below. %. This span  usually consists of active optical network elements, thus can have significant latency depending on the route and the traffic load at the midhaul network. 
 	%even though the fronthaul transport latency can largely be reduced using our proposed MESH-PON with virtualization\cite{MESH-PON}. 
 	\begin{figure*}[h]
 	    \centering
    	\includegraphics[clip, trim={0, 0, 0, 0}, width=0.87\linewidth]{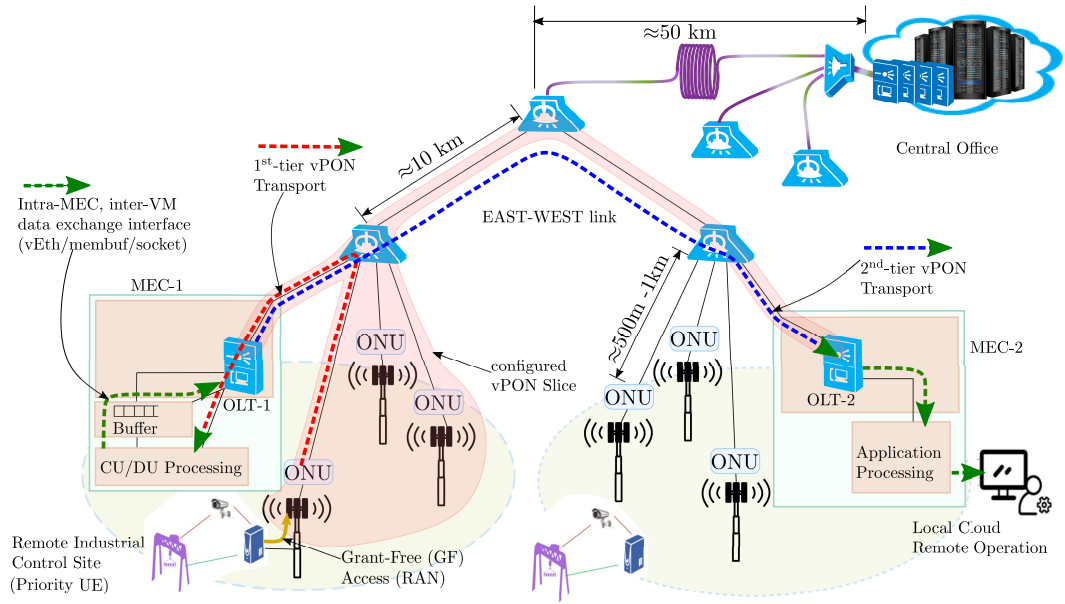}
    	\caption{System architecture and use case for the proposed scheduler's synchronization over virtualized MESH-PON}
    	\label{fig:SystemArc}
    \end{figure*}
	
    We base our architecture on virtualised PONs (vPONs) operating over a mesh access topology \cite{MESH-Networks}. A MESH-PON makes use of technology such as wavelength reflectors at splitter locations as in \cite{MESH-JOCN} (or other configurations as in \cite{Wong,Pfeiffer}) to enable direct communications between end points (i.e., without the need of OEO conversion and packet scheduling by the OLT located at the source of the PON tree in the central office). It should be noticed that a standard PON that does not support mesh connectivity can only operate as a point to multipoint. Thus the only option to achieve connectivity between end points is to communicate to an OLT located at the source of the PON tree (i.e., at the CO), thus accumulating considerable latency over each round trip. In our MESH-PON, a vPON can be dynamically created among a set of nodes that require direct communications. For example a number of small cell RUs can create a vPON that includes an MEC node that hosts the DU/CU servers controlling the small cell RUs. The virtualisation aspect (combined with a flexible and tunable physical layer) enables the connectivity among this set of nodes to be created and modified dynamically (i.e., if due to a change in load or services a set of small cell RUs needs to connect to a different MEC node). We thus propose a method to coordinate transmission from RAN to a first MEC node hosting the DU/CU (first tier) and then from there to another MEC node (second tier) hosting the application. %, over a PON operating a mesh topology (i.e., capable of connecting multiple RUs, DU/CUs and MEC nodes). 
    It should be noted that it is possible for the same MEC node to host both DU/CU and applications. However, our solution allows for a more general case, where a functional chain might be spread across more than one location. A simple reason can be due to different ownership: the owner of the C-RAN and that of the application can be different entities that run their services from different MEC nodes.  %two-tier joint virtual PON (vPON) transport method to achieve ultra-low application-level latency in MESH-PON enabled MEC based Cloud-RAN. 
    Our contribution can be summarised as follows:
    \begin{enumerate}
 		\item In the first tier, we propose an enhanced cooperative DBA which can support CGS, to achieve ultra-low latency both in the RAN and fronthaul PON transport.
 		\item In the second tier we propose an uplink-to-downlink switching mechanism between virtual PONs that minimizes latency towards the application MEC node. %vPON transports the DU processed traffic directly to the application (hosted at a different MEC) without the need transporting the traffic via OLT backplane and midhaul network. Therefore achieving an overall ultra-low end-to-end latency at the application level.
 	\end{enumerate}  
    %\vspace{-0.3cm}
    
\section{System Architecture} \label{sec:SystemModel}
	%\vspace{-0.2in}
	Fig. \ref{fig:SystemArc} presents the system architecture and use case. We consider a fixed-mobile converged architecture, where a mesh TWDM-PON is used for sharing C-RAN fronthaul with residential broadband users (not shown in the figure). For the low-latency RAN we target types of URLLC applications with latency requirements of the order of 1ms \cite{5G-NGMN-Verticals}. In order to meet this tight end-to-end latency, we propose the following coordinated two-tier vPON scheduling method.
	
	The architecture is composed of a number of PON end points serving small cell RU sites, which are provided with tunable ONU capability, and MEC nodes that are provided with tunable OLT capability (i.e., able to communicate with multiple RUs and other OLTs at the same time). Other residential users (not shown here) can be served by regular, low-cost, non-tunable ONUs. %  points 	In order to achieve the ultra-low latency in the fronthaul part of the system, the TWDM-PON architecture can be enhanced with our previously proposed virtualized MESH-PON architecture . Further, MEC nodes with limited processing capacity can be deployed in the macrocell site, where the CU/DU processing can be virtualized (vCU/vDU). 
	ONUs connected to RUs providing URLLC services (referred to as priority-UE traffic) can create a virtual PON slice with the OLT located at nearby MEC nodes (MEC-1 in this case) to transport the fronthaul data with low latency. We refer to this vPON slice as the $1^{st}$-tier and its path is illustrated with the red-colored dashed line in Fig. \ref{fig:SystemArc}.
	
	In order to achieve low latency, the first tier requires a novel, enhanced Co-DBA mechanism, to efficiently incorporate the CGS resources that can deliver URLLC. As the Radio Resource Control (RRC) block in the CU semi-statically allocates a set of CGS resources to a specific UE for URLLC services, our algorithm passes this information to the OLT, so that it can calculate uplink grants for the Co-DBA. 
	%As priority-UE's application has SLA for low-latency, the RRC module of 5G-NR semi-statically allocates a set of CGS resources which UE can acquire to send uplink traffic whenever it has application level packet to be transported. On the other hand, RRC also passes the information of the allocated CGS resources to the OLT to incorporate this CGS information while calculating uplink grants for CO-DBA. 
	A conservative approach, implemented in this work, is to consider the allocated CGS resources regardless of how many PRBs the UE is actually using. However, efficiency could be further improved by measuring the current UE traffic and then estimate the percentage of the CGS resources that is occupied in the uplink and use it along with the typical mac scheduling information for calculating the grants. %This approach allows the use of a Co-DBA mechanism also for URLLC applications, thus enabling the use of eCPRI PONs for low-latency services. %sUse of CGS resource for the uplink transmission achieves ultra-low RAN access latency while the enhanced CO-DBA ensures the corresponding uplink grants to be properly allocated by incorporating the CGS resources in order to avoid queuing of fronthaul packets due to variation of the fronthaul rates due to the UE's traffic carried in the CGS resources.
	
	The second-tier vPON slice in our proposed architecture targets ultra-low latency in the connection between the DU/CU and the application MEC nodes (shown as MEC-2 in Fig. \ref{fig:SystemArc}).   %Ideally in a typical PON based fronthaul/midhaul/backhaul deployment, in order to transport the CU/DU processed data from MEC-1 to application at MEC-2, the data has to travel all the way to CO, OEO conversion and back to MEC-2 via OLT-2. 
	%The mesh PON capability allows transport of this data directly over the mesh PON link \cite{5G-NGMN-Verticals} using our proposed second-tier vPON transport. 
	This can be done by configuring the vPON slice of OLT-1 (at MEC-1) to include the OLT at the MEC-2 as if it were an ONU and send the traffic to MEC-2 over the next downlink period of the same vPON slice. It should be noted that here we assume a worst-case scenario, where the packet waits for the next downlink frame to start, although this could be further optimised in future work. Its path is illustrated with blue-colored dashed line in Fig. \ref{fig:SystemArc}. In this case an uplink transmission (in  the $1^{st}$-tier) is followed by a downlink transmission (in the $2^{nd}$-tier), thus the packet does not need to wait for a DBA scheduling round or implement a complex inter-PON Co-DBA coordination.  %  \textcolor{red}{here we need to say how the uplink-downlink coordination is achieved}. %We refer to this as the  as UL-DL (Uplink-Downlink) method.

\section{Simulation and Results} \label{sec:Results}
The use case presented above was simulated using OMNET++. The base architecture for the simulation setup follows a MESH-PON framework (for example as described in detail in \cite{MESH-Networks}), with fibre distances reported in Fig. \ref{fig:SystemArc} and with the following enhancements. On the wireless side, two user traffic arrival processes (normal and URLLC) are created following different Poisson processes. The \ac{CGS} is abstracted in the wireless side as a set of PRBs that are statically pre-allocated whenever URLLC traffic arrives. Here, each RU uses 4 MIMO antennas and a 7.2 split, operating over 100MHz of bandwidth, where 10\% of the PRBs are semi-statically allocated/reserved as CGS resources that UEs can acquire for transmitting immediately. The transport architecture implements a MESH-PON with two tier vPON transport, with the enhanced Co-DBA proposed in this paper and the uplink-to-downlink switching scheme. The DU/CU processing time can vary depending on many factors. In order to include this type of latency, in this work we have allowed a time that is equal to the slot time (making the assumption that processing time is proportional to time slot, as shorter solt duration poses shorter timing window for the CU/DU stack processing \cite{DU_proc_Timing}).%The DU processing and the application processing timings are abstracted in the simulation framework. \textcolor{red}{what you mean by abstracted?}

\begin{figure}[h]
	\centering
	\vspace{-4mm}
	\includegraphics[clip, trim={0 0 0 0}, width=\linewidth]{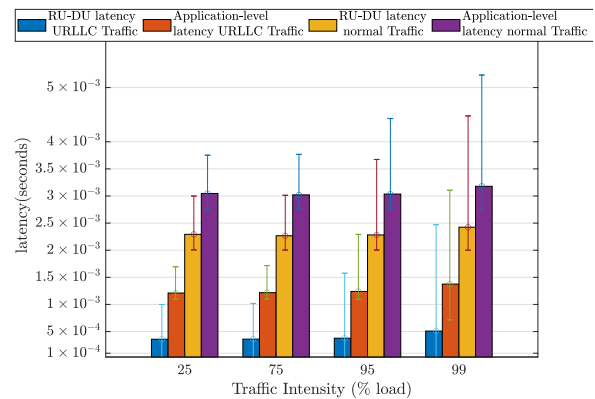}
	\caption{Average and Max latency for different traffic types and PON load (slot time =0.5ms).}
	\label{Fig:tow-tiervPON}
\end{figure}

Fig. \ref{Fig:tow-tiervPON} shows the end-to-end latency, measured both at RU-DU interface and at the application level (i.e., end-to-end), of our proposed mechanism, against the traffic load on the PON. This figure also illustrates the end-to-end latency difference between the URLLC traffic (where we employ the two-tier vPON scheme) and the normal traffic (which uses ordinary RAN scheduling, Co-DBA and a main OLT at a CO that is 50 km away). As can be seen, our proposed scheme can achieve end-to-end (application-level) average latency just above 1 ms (red bar), with maximum latency around 1.7 ms. This is unaffected by load until around 95\% traffic, when the maximum latency increases above 2 ms.

\begin{figure}[h]
	\centering
	\vspace{-4mm}
	\includegraphics[clip, trim={0 0 0 0}, width=\linewidth]{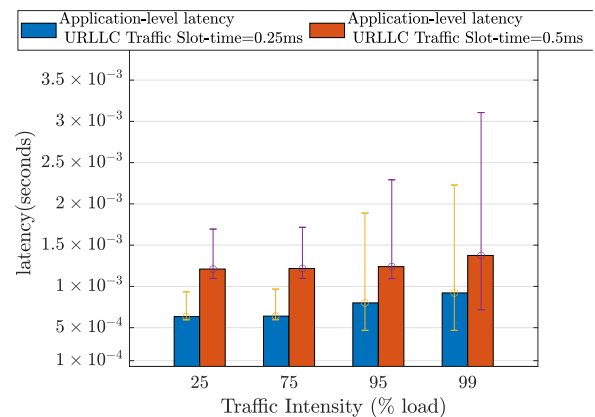}
	\caption{Average end-to-end application-level latency at different traffic load for time slot of 0.25ms and 0.5ms.}
	\label{Fig:tow-tiervPON_diffSolt}
\end{figure}

A way to further reduce latency is to reduce the RAN slot duration from 0.5 ms (used for fig. \ref{Fig:tow-tiervPON}), to 0.25 ms. This is shown in Fig. \ref{Fig:tow-tiervPON_diffSolt}. %shows the the end-to-end application level latency for the URLLC traffic (i.e, the traffic with MEC to MEC via two-tier vPON) for different NR slot configuration. 
Using a shorter NR slot configurations (for e.g, mini-slot configuration of 250 microseconds), we are able to meet sub millisecond latency (both average and max) up to 75\% load. %This is because, the DU needs to wait for the data from an entire time slot to arrive before it can start processing it. Thus a shorter time slot reduces the waiting time. %After that, it can send the DU processed data to the application at the other MEC via 2nd tier vPON. Therefore, overall latency is limited by the NR-slot duration as well. Therefore, with a shorter NR-slot configuration, our proposed scheme can achieve sub-millisecond application-level latency.

\vspace{-2mm}
\section{Conclusions} \label{sec:Conclusion}
In this paper we have shown how the use of a virtualised MESH-PON with coordinated scheduling can be instrumental in supporting URLLC latency requirements below 1 ms, on a network topology covering distances up to 20 km. The use of MESH-PON is key to support low-cost connectivity to enable densification of small cells and MEC nodes, which is a key factor for delivering the high capacity, reliability and coverage required by beyond 5G networks and applications.

\vspace{-2mm}
\section*{Acknowledgment} \label{sec:Acknowledgment}
	Financial and technical support from Intel Shannon and financial support from SFI 17/CDA/4760 and 13/RC/2077\_P2 is gratefully acknowledged.


\begin{thebibliography}{99}
	\bibitem{5G-NGMN-Verticals} Verticals URLLC Use Cases and Requirements. NGMN Allaince, Feb 2020.

	\bibitem{5G-NR_Dahlman-Scheduling}E. Dahlman, S. Parkvall, and J. Sk\"{o}ld. 5G NR: the next generation wireless access technology, CHAPTER 14: Shceduling. Elsevier, Academic Press, 2021.

	\bibitem{5G-NR_Dahlman}E. Dahlman, S. Parkvall, and J. Sk\"{o}ld. 5G NR: the next generation wireless access technology, CHAPTER 20: Industrial IoT and URLLC Enhancements. Elsevier, Academic Press, 2021.

	\bibitem{ORAN_FH_Standard} O-RAN Fronthaul Control, User and Synchronization Plane Specification 8.0. In Specification WG4: Open Fronthaul Interfaces Workgroup, Mar 2022.

    \bibitem{Co-DBA}T. Tashiro, S. Kuwano, J. Terada, T. Kawamura, N. Tanaka, S. Shigematsu, and N. Yoshimoto. A novel DBA scheme for TDM-PON based mobile fronthaul. Tu3F.3, OFC 2014.
    
	\bibitem{O-RAN-CTI} O-RAN Fronthaul Working Group 4. Cooperative Transport Interface Transport Control Plane Specification. O-RAN alliance, Mar. 2021.

	\bibitem{MESH-Networks}S. Das, F. Slyne and M. Ruffini. Optimal Slicing of Virtualised Passive Optical Networks to Support Dense Deployment of Cloud-RAN and Multi-Access Edge Computing. IEEE Network (in press), Mar 2022, arXiv preprint arXiv:2203.11857.
	
	\bibitem{MESH-JOCN}S. Das, F. Slyne, A. Kaszubowska and M. Ruffini. Virtualised EAST-WEST PON Architecture Supporting Low-Latency communication for Mobile Functional-Split Based on Multi-Access Edge Computing. JOCN, 10(12), Oct. 2020

    \bibitem{Wong} C. Ranaweera, E. Wong, C. Lim, and A. Nirmalathas. Next generation optical-wireless converged network architectures. IEEE Network 26, 22-27 (2012).
    
    \bibitem{Pfeiffer} T. Pfeiffer. Converged heterogeneous optical metro-access networks. Tu.5.B.1, OFC 2010.
    
	\bibitem{DU_proc_Timing} J. S. Panchal, S. Subramanian and R. Cavatur, Enabling and Scaling of URLLC Verticals on 5G vRAN Running on COTS Hardware, in IEEE Communications Magazine, vol. 59, no. 9, pp. 105-111, September 2021

\end{thebibliography}
\end{document}